\documentclass[12pt] {article}
\usepackage{psfrag}
\usepackage{graphicx}
\usepackage{latexsym,amsfonts}
\usepackage{amssymb}
\begin{document}
\setcounter{page}{1}
\def\theequation{\arabic{section}.\arabic{equation}}
\def\theequation{\thesection.\arabic{equation}}
\setcounter{section}{0}

\title{Dynamical breaking of conformal symmetry in the massless 
Thirring model}

\author{M. Faber\thanks{E--mail: faber@kph.tuwien.ac.at, Tel.:
+43--1--58801--14261, Fax: +43--1--58801--14299} ~~and~
A. N. Ivanov\thanks{E--mail: ivanov@kph.tuwien.ac.at, Tel.:
+43--1--58801--14261, Fax: +43--1--58801--14299}~\thanks{Permanent
Address: State Polytechnical University, Department of Nuclear
Physics, 195251 St. Petersburg, Russian Federation}}

\date{\today}

\maketitle

\vspace{-0.5in}
\begin{center}
{\it Atominstitut der \"Osterreichischen Universit\"aten,
Arbeitsbereich Kernphysik und Nukleare Astrophysik, Technische
Universit\"at Wien, \\ Wiedner Hauptstr. 8-10, A-1040 Wien,
\"Osterreich }
\end{center}

\begin{center}
\begin{abstract}
We discuss conformal invariance of the massless Thirring model. We
show that conformal symmetry of the massless Thirring model is
dynamically broken due to the constant of motion caused by the
equations of motion.  This confirms the existence of the chirally
broken phase in the massless Thirring model (Eur. Phys. J. C {\bf 20},
723 (2001), which is accompanied by the appearance of massless
(pseudo)scalar Goldstone bosons (Eur. Phys. J. C {\bf 24}, 653 (2002),
hep--th/0210104 and hep--th/0305174).
\end{abstract}

PACS: 02.30.Ik, 11.10.Cd, 11.10.Ef, 11.10.Kk
\end{center}

\newpage

\section{Introduction}
\setcounter{equation}{0}

\hspace{0.2in} The massless Thirring model \cite{WT58} is a theory of
a self--coupled Dirac field $\psi(x)$
\begin{eqnarray}\label{label1.1}
{\cal L}_{\rm Th}(x) = \bar{\psi}(x)i\gamma^{\mu}\partial_{\mu}\psi(x) -
\frac{1}{2}\,g\,\bar{\psi}(x)\gamma^{\mu}\psi(x)\bar{\psi}(x)
\gamma_{\mu}\psi(x),
\end{eqnarray}
where $g$ is a dimensionless coupling constant that can be both
positive and negative as well. The field $\psi(x)$ is a spinor field
with two components $\psi_1(x)$ and $\psi_2(x)$, $x$ is a 2--vector
$x^{\mu} = (x^0, x^1)$, where $x^0$ and $x^1$ are time and spatial
components. The $\gamma$--matrices are defined by $\gamma^0 =
\sigma_1$, $\gamma^1 = -i\,\sigma_1$ and $\gamma^5 = \gamma^0\gamma^1
= \sigma_3$, where $\sigma_i\,(i = 1,2,3)$ are $2\times 2$ Pauli
matrices.  These $\gamma$--matrices obey the relations \cite{FI1}
\begin{eqnarray}\label{label1.2}
\gamma^{\mu}\gamma^{\nu}&+& \gamma^{\nu}\gamma^{\mu} = 2
g^{\mu\nu}\quad,\quad \gamma^{\mu}\gamma^5 + \gamma^5\gamma^{\mu} =
0,\nonumber\\ \nonumber\\ \gamma^{\mu}\gamma^{\nu} &=& g^{\mu\nu} +
\varepsilon^{\mu\nu}\gamma^5 \quad,\quad \gamma^{\mu}\gamma^5 = -
\varepsilon^{\mu\nu}\gamma_{\nu},
\end{eqnarray}
where $\varepsilon^{\mu\nu}$ is the Levy--Civita tensor defined by
$\varepsilon^{01} = - \varepsilon_{01} = 1$. The Lagrangian
(\ref{label1.1}) is invariant under the chiral group $U_{\rm
V}(1)\times U_{\rm A}(1)$
\begin{eqnarray}\label{label1.3}
\psi(x) \stackrel{\rm V}{\longrightarrow} \psi^{\prime}(x) &=&
e^{\textstyle i\alpha_{\rm V}}\psi(x) ,\nonumber\\ \psi(x)
\stackrel{\rm A}{\longrightarrow} \psi^{\prime}(x) &=& e^{\textstyle
i\alpha_{\rm A}\gamma^5}\psi(x),
\end{eqnarray}
where $\alpha_{\rm V}$ and $\alpha_{\rm A}$ are real parameters
defining global rotations. Due to invariance under the chiral group
$U_{\rm V}(1)\times U_{\rm A}(1)$ the vector and axial--vector current
$j^{\mu}(x)$ and $j^{\mu}_5(x)$, induced by vector (V) and
axial--vector (A) rotations and defined by
\begin{eqnarray}\label{label1.4}
j^{\mu}(x) &=& \bar{\psi}(x)\gamma^{\mu}\psi(x),\nonumber\\
j^{\mu}_5(x) &=& \bar{\psi}(x)\gamma^{\mu}\gamma^5\psi(x),
\end{eqnarray}
are conserved $\partial_{\mu}j^{\mu}(x) = \partial_{\mu}j^{\mu}_5(x) =
0$.  Recall, that in 1+1--dimensional field theories the vector and
axial--vector currents are related by $j^{\mu}_5(x) = -
\varepsilon^{\mu\nu}j_{\nu}(x)$ due to the properties of Dirac
matrices.

In addition to chiral invariance the massless Thirring model is
invariant under the conformal group \cite{LH79,EF96}, which contains
the Poincar${\acute{e}}$ group supplemented by dilatations $x_{\mu}
\to x_{\mu}' = x_{\mu}/\rho$ and inversions $x_{\mu} \to x_{\mu}' =
c\,x_{\mu}/x^2$, where $\rho$ and $c$ are parameters of the
transformations \cite{EF96}--\cite{CHP}.

If the conformal invariance of the massless Thirring model can be
broken neither explicitly nor spontaneously, the spontaneous breaking
of chiral symmetry of the massless Thirring model found in
\cite{FI1,FI2}, leading to the breakdown of conformal symmetry of the
massless Thirring model, becomes never possible.

Below we show that conformal symmetry of the massless Thirring model
becomes dynamically broken due to the constant of motion following
from the equations of motion for the massless Thirring fermion fields
\cite{FI1}. This allows the existence of the chirally broken phase in
the massless Thirring model obtained in \cite{FI1,FI2}. Since
dilatations are a part of the conformal group, the dynamical breaking
of dilatational invariance of the massless Thirring model should
testify the dynamical breaking of a conformal symmetry.

Under dilatations $x_{\mu} \to x_{\mu}' = x_{\mu}/\rho$ the massless
Thirring fermion fields behave as follows \cite{KW70} (see also
\cite{CHP})
\begin{eqnarray}\label{label1.5}
U(\rho)\,\psi(x)\,U^{\dagger}(\rho) = \rho^{\;d}\,\psi(\rho x),
\end{eqnarray}
where $d$ is a dimension of the field $\psi(x)$. A unitary operator
$U(\rho)$ is defined by \cite{CHP}
\begin{eqnarray}\label{label1.6}
U(\rho) = e^{\textstyle\,i\,{\ell n}\rho\,D},
\end{eqnarray}
where $D$ is the generator of the scale transformations (dilatations).
 For the infinitesimal dilatation $\rho = 1 + \epsilon$ the operator
 $D$ satisfies the commutation relation
\begin{eqnarray}\label{label1.7}
i[D,\psi(x)] = (d + x^{\mu}\partial_{\mu})\psi(x).
\end{eqnarray}
Under scale transformation (\ref{label1.5}) the Lagrangian
(\ref{label1.1}) transforms as follows
\begin{eqnarray}\label{label1.8}
{\cal L}_{\rm Th}(x) \to {\cal L}_{\rm Th}(\rho x) = \rho^{\;2d +
1}\bar{\psi}(\rho x)i\gamma^{\mu}\frac{\partial \psi(\rho x)}{\partial
(\rho x)^{\mu}} - \rho^{\;4d}\,\frac{1}{2}\,g\, \bar{\psi}(\rho
x)\gamma^{\mu}\psi(\rho x)\bar{\psi}(\rho x)\gamma_{\mu}\psi(\rho x).
\end{eqnarray}
The action ${\cal S}_{\rm Th}[\psi,\bar{\psi}]$ we determine by
\begin{eqnarray}\label{label1.9}
{\cal S}_{\rm Th}[\psi,\bar{\psi}] = \int d^2x\,{\cal L}_{\rm Th}(x).
\end{eqnarray}
Due to the scale transformation (\ref{label1.5}) it changes as
\begin{eqnarray}\label{label1.10}
\hspace{-0.3in}{\cal S}_{\rm Th}[\psi,\bar{\psi}] &=& \int d^2x\,{\cal
L}_{\rm Th}(x) \to \rho^{2d - 1}\int d^2(\rho x)\,\Big(\bar{\psi}(\rho
x)i\gamma^{\mu}\frac{\partial \psi(\rho x)}{\partial (\rho
x)^{\mu}}\nonumber\\ \hspace{-0.3in}&& - \rho^{\;2d -
1}\,\frac{1}{2}\,g\, \bar{\psi}(\rho x)\gamma^{\mu}\psi(\rho
x)\bar{\psi}(\rho x) \gamma_{\mu}\psi(\rho x)\Big).
\end{eqnarray}
It is seen that the action is invariant under the scale
transformations (\ref{label1.5}) if the dimension of the massless
Thirring fermion field is equal to $d = 1/2$. This is the so--called
canonical dimension of a fermion field in 1+1--dimensional
space--time.

For $d = 1/2$ the Lagrangian ${\cal L}_{\rm Th}(\rho x)$ differs from
the Lagrangian ${\cal L}_{\rm Th}(x)$ by a constant factor
$\rho^2$. This means that the equations of motion of the massless
Thirring fermion fields should be invariant under scale
transformations.

As has been shown in \cite{FI1}, the equations of motion for the
massless Thirring fermion fields lead to a constant of motion, which
is not invariant under scale transformations $x_{\mu} \to x_{\mu}' =
x_{\mu}/\rho$ and (\ref{label1.5}). This should testify a dynamical
breaking of the dilatational invariance of the Thirring model.

The paper is organized as follows. In Section 2 we analyse the
equations of motion for classical massless Thirring fermion fields and
derive the constant of motion. We show that the constant of motion
breaks dynamically dilatational and conformal symmetries. In Section 3
we derive the equations of motion for the quantum massless Thirring
fermion fields and the quantum version of the constant of motion
breaking dilatational and conformal symmetries in the quantum massless
Thirring model. In the Conclusion we discuss the obtained results in
connection with the existence of the chirally broken phase of the
massless Thirring model pointed out in \cite{FI1,FI2}.

\section{Constant of motion and dynamical breaking of dilatational 
invariance. Classical fermion fields}
\setcounter{equation}{0}

\hspace{0.2in} In this section we deal with classical Thirring fermion
fields and show that the classical equations of motion lead to the
constant of motion found in \cite{FI1}. For the analysis of the
evolution of the classical Thirring fermion fields the Lagrangian
(\ref{label1.1}) does not need to be taken in the normal--ordered form
\cite{FI1}. The equations of motion read \cite{FI1}
\begin{eqnarray}\label{label2.1}
i\gamma^{\mu}\partial_{\mu}\psi(x) &=&
g\,j^{\mu}(x)\gamma_{\mu}\psi(x),\nonumber\\
-i\partial_{\mu}\bar{\psi}(x)\gamma^{\mu}&=&
g\,\bar{\psi}(x)\gamma_{\mu}j^{\mu}(x).
\end{eqnarray}
Due to the peculiar properties of 1+1--dimensional quantum field
theories of fermion fields \cite{CS68} the equations of motion
(\ref{label2.1}) are equivalent to \cite{FI1}
\begin{eqnarray}\label{label2.2}
i\partial_{\mu}\psi(x) &=& a\,j_{\mu}(x)\,\psi(x) +
b\,\varepsilon_{\mu\nu}\,j^{\nu}(x)\gamma^5\psi(x),\nonumber\\
-i\partial_{\mu}\bar{\psi}(x) &=& a\,\bar{\psi}(x)\,j_{\mu}(x) +
b\,\bar{\psi}(x)\gamma^5j^{\nu}(x)\,\varepsilon_{\nu\mu},
\end{eqnarray}
where the parameters $a$ and $b$ are equal to \cite{FI1}
\begin{eqnarray}\label{label2.3}
a &=& \frac{1}{2}\,\Big(g + \frac{1}{c}\Big),\nonumber\\
b &=& \frac{1}{2}\,\Big(g - \frac{1}{c}\Big),
\end{eqnarray}
where $c$ is the Schwinger term \cite{FI1}.

The equations of motion (\ref{label2.2}) can be transcribed into the
equations of motion for the scalar and pseudoscalar fermion densities
$\bar{\psi}(x)\psi(x)$ and $\bar{\psi}(x) i \gamma^5 \psi(x)$. They
read
\begin{eqnarray}\label{label2.4}
\partial_{\mu}[\bar{\psi}(x)\psi(x)] &=& -
2b\,\varepsilon_{\mu\nu}j^{\nu}(x)\,[\bar{\psi}(x) i \gamma^5
\psi(x)],\nonumber\\ \partial_{\mu}[\bar{\psi}(x)i\gamma^5\psi(x)] &=&
+ 2b\,\varepsilon_{\mu\nu}j^{\nu}(x)\,[\bar{\psi}(x) \psi(x)].
\end{eqnarray}
Multiplying the first equation by $\bar{\psi}(x)\psi(x)$ and the second
by $\bar{\psi}(x) i \gamma^5 \psi(x)$ and summing up the obtained
expressions we arrive at the relation \cite{FI1}
\begin{eqnarray}\label{label2.5}
\frac{\partial}{\partial x^{\mu}}\Big([\bar{\psi}(x)\psi(x)]^2 +
[\bar{\psi}(x)i\gamma^5\psi(x)]^2\Big) = 0.
\end{eqnarray}
Hence, the expression in the parentheses is the constant of motion
\begin{eqnarray}\label{label2.6}
[\bar{\psi}(x)\psi(x)]^2 +
[\bar{\psi}(x)i\gamma^5\psi(x)]^2 = C,
\end{eqnarray}
where $C$ is a constant. As has been shown in \cite{FI1}, this is equal
to $C = M^2/g^2$, where $M$ is the dynamical mass of Thirring fermion
fields.

Applying a Fierz transformation and multiplying both sides by $g/2$ we
transform the constant of motion (\ref{label2.6}) to the form
\cite{FI1}
\begin{eqnarray}\label{label2.7}
\frac{1}{2}\,g\,\bar{\psi}(x)\gamma^{\mu}\psi(x)\bar{\psi}(x)
\gamma_{\mu}\psi(x) = - \frac{M^2}{2g~} = \frac{2\pi}{g}\,{\cal E}[M],
\end{eqnarray}
where ${\cal E}[M] = - M^2/4\pi$ is the minimum of the energy density
of the ground state of the Thirring fermion fields in the chirally
broken phase \cite{FI1}.

Since the l.h.s. of (\ref{label2.7}) is a potential of the
self--coupled Thirring fermions, one can conclude that relation
(\ref{label2.7}) testifies the evolution of the Thirring fermions with
a constant potential energy in the proximity of the minimum of the
energy density of the ground state of the massless Thirring model in
the chirally broken phase.

Under scale transformation the l.h.s. of (\ref{label2.6}) changes as
follows
\begin{eqnarray}\label{label2.8}
\rho^{\;4d}\,([\bar{\psi}(\rho x)\psi(\rho x)]^2 + [\bar{\psi}(\rho
x)i\gamma^5\psi(\rho x)]^2) = C.
\end{eqnarray}
Since $[\bar{\psi}(x)\psi(x)]^2 + [\bar{\psi}(x)i\gamma^5\psi(x)]^2$
is a constant of motion, it is obvious that
\begin{eqnarray}\label{label2.9}
[\bar{\psi}(\rho x)\psi(\rho x)]^2 + [\bar{\psi}(\rho
x)i\gamma^5\psi(\rho x)]^2 = [\bar{\psi}(x)\psi(x)]^2 +
[\bar{\psi}(x)i\gamma^5\psi(x)]^2 = C.
\end{eqnarray}
Substituting (\ref{label2.9}) in (\ref{label2.8}) we obtain the
relation
\begin{eqnarray}\label{label2.10}
\rho^{\;4d}\,C = C
\end{eqnarray}
which is obviously broken for $\rho \neq 1$ for $C \neq 0$, even if
the dimension $d$ of the Thirring fermion fields is equal to the
canonical dimension $d = 1/2$.

The constraint (\ref{label2.10}) is valid only for $C = 0$. However,
the vanishing constant of motion entails the trivial solution of the
equations of motion of the massless Thirring model, $\psi(x) =
0$. Hence, for any non--trivial solution dilatational symmetry as well
as conformal symmetry becomes dynamically broken due to the constant
of motion.

This means that the massless Thirring fermion fields evolve conserving
the constant of motion and breaking scale invariance. Dynamical
breaking of the scale invariance is equivalent to the dynamical
breaking of conformal symmetry \cite{RJ71}. This agrees with the
assertion by Fradkin and Palchik \cite{EF96} that conformal symmetry
in the massless Thirring model can be spontaneously broken.

\section{Constant of motion and dynamical breaking of dilatational 
invariance. Quantum fermion fields}
\setcounter{equation}{0}

\hspace{0.2in} In this section we analyse the existence of the
constant of motion (\ref{label2.6}) for quantum massless Thirring
fermion fields. Since the constant of motion is the consequence of the
equations of motion, we have to derive the quantum equations of
motion. For the derivation of quantum equations of motion the
Lagrangian (\ref{label1.1}) should be taken in the normal--ordered
form \cite{FI1}
\begin{eqnarray}\label{label3.1}
{\cal L}_{\rm Th}(x) =
:\bar{\psi}(x)i\gamma^{\mu}\partial_{\mu}\psi(x): -
\frac{1}{2}\,g\,:\bar{\psi}(x)\gamma^{\mu}\psi(x)\bar{\psi}(x)
\gamma_{\mu}\psi(x):,
\end{eqnarray}
where $:\ldots:$ indicates  the normal ordering.

For further transformations we decompose conventionally Thirring
fermion fields $\psi(x)$ and $\bar{\psi}(x)$ into positive and
negative frequency parts \cite{GW50}
\begin{eqnarray}\label{label3.2}
\psi(x) &=& \psi^{(+)}(x) + \psi^{(-)}(x),\nonumber\\ \bar{\psi}(x)
 &=& \bar{\psi}^{(+)}(x) + \bar{\psi}^{(-)}(x),
\end{eqnarray}
Due to non--linearity the operators
$\psi^{(+)}(x)\,(\bar{\psi}^{(-)}(x))$ and
$\psi^{(-)}(x)\,(\bar{\psi}^{(+)}(x))$ do not annihilate a fermion
(anti--fermion) or create of an anti--fermion (fermion). However, the
normal--ordering assumes that all operators
$\psi^{(-)}(x)\,(\bar{\psi}^{(+)}(x))$ should stand to left from the
operators $\psi^{(+)}(x)\,(\bar{\psi}^{(-)}(x))$.

In terms of the positive and negative frequency parts of the fermion
fields $\psi(x)$ and $\bar{\psi}(x)$ the Lagrangian (\ref{label3.1})
acquires the form
\begin{eqnarray}\label{label3.3}
\hspace{-0.3in}{\cal L}_{\rm Th}(x) &=& \bar{\psi}^{(+)}_a(x)
(i\gamma^{\mu})_{ab} \partial_{\mu} \psi^{(+)}_b(x) -
(i\gamma^{\mu})_{ab} \partial_{\mu}
\psi^{(-)}_b(x)\bar{\psi}^{(-)}_a(x)\nonumber\\
\hspace{-0.3in}&+& \bar{\psi}^{(+)}_a(x) (i\gamma^{\mu})_{ab}
\partial_{\mu} \psi^{(-)}_b(x) +
\bar{\psi}^{(-)}_a(x)(i\gamma^{\mu})_{ab} \partial_{\mu}
\psi^{(+)}_b(x)\nonumber\\
\hspace{-0.3in}&-&
\frac{1}{2}\,g\,(\gamma^{\mu})_{ab}(\gamma_{\mu})_{cd}\,\{-
\bar{\psi}^{(+)}_a(x) \bar{\psi}^{(+)}_c(x) \psi^{(+)}_b(x)
\psi^{(+)}_d(x)\nonumber\\
\hspace{-0.3in}&+& \bar{\psi}^{(+)}_c(x) \psi^{(-)}_b(x)
\psi^{(+)}_d(x) \bar{\psi}^{(-)}_a(x) + \bar{\psi}^{(+)}_c(x)
\bar{\psi}^{(+)}_a(x) \psi^{(-)}_b(x) \psi^{(+)}_d(x)\nonumber\\
\hspace{-0.3in}&-& \bar{\psi}^{(+)}_c(x) \psi^{(+)}_d(x)
\psi^{(+)}_b(x) \bar{\psi}^{(-)}_a(x) +
\bar{\psi}^{(+)}_a(x)\psi^{(-)}_d(x)\psi^{(+)}_b(x)
\bar{\psi}^{(-)}_c(x)\nonumber\\
\hspace{-0.3in}&-& \psi^{(-)}_b(x) \psi^{(-)}_d(x)
\bar{\psi}^{(-)}_a(x) \bar{\psi}^{(-)}_c(x) - \bar{\psi}^{(+)}_a(x)
\psi^{(-)}_b(x) \psi^{(-)}_d(x) \bar{\psi}^{(-)}_c(x)\nonumber\\
\hspace{-0.3in}&+& \psi^{(-)}_d(x) \psi^{(+)}_b(x)
\bar{\psi}^{(-)}_a(x) \bar{\psi}^{(-)}_c(x) + \bar{\psi}^{(+)}_a(x)
\bar{\psi}^{(+)}_c(x)\psi^{(-)}_d(x) \psi^{(+)}_b(x)\nonumber\\
\hspace{-0.3in}&-& \psi^{(-)}_b(x) \bar{\psi}^{(+)}_c(x)
\psi^{(-)}_d(x) \bar{\psi}^{(-)}_a(x) + \bar{\psi}^{(+)}_a(x)
\psi^{(-)}_b(x) \bar{\psi}^{(+)}_c(x) \psi^{(-)}_d(x) \nonumber\\
\hspace{-0.3in}&-& \bar{\psi}^{(+)}_c(x) \psi^{(-)}_d(x)
\psi^{(+)}_b(x) \bar{\psi}^{(-)}_a(x) -
\bar{\psi}^{(+)}_a(x)\psi^{(+)}_b(x) \psi^{(+)}_d(x)
\bar{\psi}^{(-)}_c(x)\nonumber\\
\hspace{-0.3in}&+& \psi^{(-)}_b(x) \bar{\psi}^{(-)}_a(x)
\psi^{(+)}_d(x) \bar{\psi}^{(-)}_c(x) - \bar{\psi}^{(+)}_a(x)
\psi^{(-)}_b(x) \psi^{(+)}_d(x) \bar{\psi}^{(-)}_c(x)\nonumber\\
\hspace{-0.3in}&+& \psi^{(+)}_b(x) \bar{\psi}^{(-)}_a(x)
\psi^{(+)}_d(x) \bar{\psi}^{(-)}_c(x)\}.
\end{eqnarray}
The quantum equations of motion can be obtained by differentiating the
Lagrangian (\ref{label3.3}) with respect to
$\bar{\psi}^{(+)}_a(x)$. This gives
\begin{eqnarray}\label{label3.4}
\hspace{-0.3in}&&i(\gamma^{\mu})_{ab}\partial_{\mu}\psi^{(+)}_b(x) +
i(\gamma^{\mu})_{ab}\partial_{\mu}\psi^{(-)}_b(x) =
\frac{1}{2}\,g\,(\gamma^{\mu})_{ab}(\gamma_{\mu})_{cd}\{-
\bar{\psi}^{(+)}_c(x) \psi^{(+)}_b(x) \psi^{(+)}_d(x)\nonumber\\
\hspace{-0.3in}&& - \bar{\psi}^{(+)}_c(x) \psi^{(-)}_b(x)
\psi^{(+)}_d(x) + \psi^{(-)}_d(x) \psi^{(+)}_b(x)
\bar{\psi}^{(-)}_c(x) - \psi^{(-)}_b(x) \psi^{(-)}_d(x)
\bar{\psi}^{(-)}_c(x)\nonumber\\
\hspace{-0.3in}&& + \bar{\psi}^{(+)}_c(x) \psi^{(-)}_d(x)
\psi^{(+)}_b(x) + \psi^{(-)}_b(x) \bar{\psi}^{(+)}_c(x)
\psi^{(-)}_d(x) - \psi^{(+)}_b(x) \psi^{(+)}_d(x)
\bar{\psi}^{(-)}_c(x)\nonumber\\
\hspace{-0.3in}&&  - \psi^{(-)}_b(x) \psi^{(+)}_d(x)
\bar{\psi}^{(-)}_c(x)  + \bar{\psi}^{(+)}_c(x) \psi^{(+)}_d(x)
\psi^{(+)}_b(x) + \bar{\psi}^{(-)}_d(x) \psi^{(+)}_b(x)
\bar{\psi}^{(-)}_c(x)\nonumber\\
\hspace{-0.3in}&& + \bar{\psi}^{(+)}_c(x) \psi^{(-)}_d(x)
\psi^{(+)}_b(x) - \psi^{(+)}_b(x) \psi^{(+)}_d(x)
\bar{\psi}^{(-)}_c(x) - \bar{\psi}^{(+)}_c(x) \psi^{(-)}_b(x)
\psi^{(+)}_d(x)\nonumber\\
\hspace{-0.3in}&& + \psi^{(-)}_d(x) \psi^{(-)}_b(x)
\bar{\psi}^{(-)}_c(x) - \psi^{(-)}_b(x) \psi^{(+)}_d(x)
\bar{\psi}^{(-)}_c(x) + \bar{\psi}^{(+)}_c(x) \psi^{(-)}_d(x)
\psi^{(-)}_b(x)\}.
\end{eqnarray}
These equations of motion can be rewritten in the usual form
\begin{eqnarray}\label{label3.5}
i\gamma^{\mu}\partial_{\mu}\psi(x) &=&
\frac{1}{2}\,g\,:\bar{\psi}(x)\gamma^{\mu}\psi(x)\gamma_{\mu}\psi(x) +
\gamma_{\mu}\psi(x)\bar{\psi}(x)\gamma^{\mu}\psi(x): =\nonumber\\
&=& g\,:\{\bar{\psi}(x)\gamma^{\mu}\psi(x),\gamma_{\mu}\psi(x)\}:.
\end{eqnarray}
For the Dirac conjugate fermion field the equations of motion read
\begin{eqnarray}\label{label3.6}
- i\partial_{\mu}\bar{\psi}(x)\gamma^{\mu} &=&
\frac{1}{2}\,g\,:\bar{\psi}(x) \gamma_{\mu} \bar{\psi}(x) \gamma^{\mu}
\psi(x) + \bar{\psi}(x) \gamma^{\mu} \psi(x)
\bar{\psi}(x)\gamma_{\mu}: = \nonumber\\ &=& g\,:\{\bar{\psi}(x)
\gamma_{\mu}, \bar{\psi}(x) \gamma^{\mu} \psi(x)\}:.
\end{eqnarray}
The quantum analogies of the equations (\ref{label2.2}) read
\begin{eqnarray}\label{label3.7}
i\partial_{\mu}\psi(x) &=&
a\,:\{\bar{\psi}(x)\gamma_{\mu}\psi(x),\psi(x)\}: +
b\,\varepsilon_{\mu\nu}\,:\{\bar{\psi}(x)\gamma^{\nu}\psi(x),\gamma^5\psi(x)\}:,\nonumber\\
- i\partial_{\mu}\bar{\psi}(x) &=& a\,:\{\bar{\psi}(x), \bar{\psi}(x)
\gamma_{\mu} \psi(x)\}: -
b\,\varepsilon_{\mu\nu}:\{\bar{\psi}(x)\gamma^5, \bar{\psi}(x)
\gamma^{\nu} \psi(x)\}:.
\end{eqnarray}
For the quantum versions of (\ref{label2.4}) we get 
\begin{eqnarray}\label{label3.8}
\hspace{-0.3in}&&\partial_{\mu}[\bar{\psi}(x)\psi(x)] = -
i\,a\,\bar{\psi}(x):\{\bar{\psi}(x)\gamma_{\mu}\psi(x),\psi(x)\}: +
i\,a:\{\bar{\psi}(x), \bar{\psi}(x) \gamma_{\mu}
\psi(x)\}:\psi(x)\nonumber\\
\hspace{-0.3in}&&-\,i\,b\,\varepsilon_{\mu\nu}\,
\bar{\psi}(x):\{\bar{\psi}(x)\gamma^{\nu}\psi(x),\gamma^5\psi(x)\}: -
i\,b\,\varepsilon_{\mu\nu}:\{\bar{\psi}(x)\gamma^5, \bar{\psi}(x)
\gamma^{\nu} \psi(x)\}:\psi(x),\nonumber\\
\hspace{-0.3in}&&\partial_{\mu}[\bar{\psi}(x) i \gamma^5 \psi(x)] =
a\,\bar{\psi}(x)\gamma^5
:\{\bar{\psi}(x)\gamma_{\mu}\psi(x),\psi(x)\}: -a\,:\{\bar{\psi}(x),
\bar{\psi}(x) \gamma_{\mu} \psi(x)\}:\gamma^5 \psi(x)\nonumber\\
\hspace{-0.3in}&&+
b\,\varepsilon_{\mu\nu}\,\bar{\psi}(x)\gamma^5:\{\bar{\psi}(x)
\gamma^{\nu}\psi(x), \gamma^5 \psi(x)\}: + b\,\varepsilon_{\mu\nu}
:\{\bar{\psi}(x)\gamma^5, \bar{\psi}(x) \gamma^{\nu}
\psi(x)\}:\gamma^5\psi(x).
\end{eqnarray}
For the transformation of the r.h.s. of (\ref{label3.8}) we would use
Wick's theorem \cite{GW50,IZ80}. This yields
\begin{eqnarray}\label{label3.9}
\hspace{-0.3in}\partial_{\mu}[\bar{\psi}(x)\psi(x)] &=& + (a +
b)\,\lim_{y\to x}:\bar{\psi}(x)i\gamma^5 \psi(x):\,{\rm
tr}\{\gamma_{\mu}\gamma^5 S_F(x - y)\},\nonumber\\
\hspace{-0.3in}\partial_{\mu}[\bar{\psi}(x)i\gamma^5\psi(x)] &=& - (a
+ b)\,\lim_{y\to x}:\bar{\psi}(x)\psi(x):\,{\rm
tr}\{\gamma_{\mu}\gamma^5 S_F(x - y)\},
\end{eqnarray}
where $S_F(x - y)$ is the exact causal two--point Green function of
the massless Thirring fermion fields which we define in the form of
the K\"allen--Lehmann representation \cite{IZ80}
\begin{eqnarray}\label{label3.10}
\hspace{-0.3in}&&S_F(x - y) = i\langle 0|{\rm
T}(\psi(x)\bar{\psi}(y))|0\rangle = \nonumber\\
\hspace{-0.3in}&&= -
\frac{1}{2\pi}\,\gamma^{\mu}\frac{\partial}{\partial
x^{\mu}}\int^{\infty}_0dm^2\,\rho(m^2)\,K_0(m\sqrt{ - (x - y)^2 +
i\,0}) =\nonumber\\ \hspace{-0.3in}&&=-
\frac{1}{2\pi}\,\frac{\gamma^{\mu}(x - y)_{\mu}}{\sqrt{ - (x - y)^2 +
i\,0}}\int^{\infty}_0dm^2\,\rho(m^2)\,m\,K_1(m\sqrt{ - (x - y)^2 +
i\,0}),
\end{eqnarray}
where $K_0(z)$ and $K_0(z)$ are McDonald's functions and $\rho(m^2)$
is the K\"allen--Lehmann spectral function.

It is obvious that the equations (\ref{label3.9}) can be written as
\begin{eqnarray}\label{label3.11}
\partial_{\mu}(:[\bar{\psi}(x)\psi(x)]^2 +
[\bar{\psi}(x)i\gamma^5\psi(x)]^2:) & =&(a + b)\,\lim_{y\to
x}[:\bar{\psi}(x)\psi(x):,:\bar{\psi}(x)i\gamma^5 \psi(x):]\nonumber\\
&&\times\,{\rm tr}\{\gamma_{\mu}\gamma^5 S_F(x - y)\}.
\end{eqnarray}
Since the scalar and pseudoscalar fermion densities commute, we arrive
at the equation
\begin{eqnarray}\label{label3.12}
\partial_{\mu}(:[\bar{\psi}(x)\psi(x)]^2 +
[\bar{\psi}(x)i\gamma^5\psi(x)]^2:) = 0.
\end{eqnarray}
Thus the quantum version of the constant of motion reads
\begin{eqnarray}\label{label3.13}
:[\bar{\psi}(x)\psi(x)]^2 + [\bar{\psi}(x)i\gamma^5\psi(x)]^2:\, =\,C,
\end{eqnarray}
where $C = M^2/g^2$. It differs from the constant of motion of the
classical equations of motion (\ref{label2.6}) only by the normal
ordering of the field operators. By a Fierz transformation we obtain
\begin{eqnarray}\label{label3.14}
:\bar{\psi}(x)\gamma_{\mu}\psi(x) \bar{\psi}(x)\gamma^{\mu}\psi(x): =
- C.
\end{eqnarray}
In the component form the constant of motion reads
\begin{eqnarray}\label{label3.15}
:\psi^{\dagger}_1(x)\psi_1(x)\psi^{\dagger}_2(x)\psi_2(x): \,=\,
\frac{1}{4}\,C.
\end{eqnarray}
This constant of motion for $C \neq 0$ breaks dynamically both
dilatational and conformal symmetry of the massless Thirring model.

It is obvious that for $\,C = 0$, that is demanded by dilatational and
conformal invariance, there can be only a trivial solution of the
quantum equations of motion, i.e.  $\psi_1(x) = \psi_2(x) = 0$. Hence,
the conformal invariant massless Thirring model does not exist.

\section{Conclusion}

\hspace{0.2in} We have discussed the constant of motion for the
evolution of the massless Thirring fermion fields in connection with
dynamical breaking of dilatational and conformal symmetries of the
massless Thirring model. We have derived the constant of motion for
both classical and quantum massless Thirring fermion field. The
existence of the constant of motion for the evolution of the massless
Thirring fermion fields has been recently confirmed within the free
massless boson field representation of the massless Thirring fermion
fields \cite{FI6}.

We have shown that the constant of motion of the massless Thirring
fermion fields breaks dynamically both dilatational and conformal
symmetries of the massless Thirring model. This agrees with Fradkin
and Palchik \cite{EF96}. The vanishing constant of motion, compatible
with dilatational and conformal invariance of the massless Thirring
model, leads to a trivial solution of the classical and quantum
equations of motion, $\psi(x) = 0$. This means that the massless
Thirring model, invariant under dilatational and conformal symmetry,
does not exist.

It is well--known from low--energy hadronic physics that spontaneous
breaking of dilatational symmetry entails spontaneous breaking of
chiral symmetry \cite{CHP}. Thus, the dynamical breaking of
dilatational and conformal symmetries of the massless Thirring model,
due to the constant of motion, testifies the validity of our results
concerning the existence of the chirally broken phase of the massless
Thirring model pointed out in \cite{FI1,FI2}.

We would like to emphasize that the dynamical breaking of conformal
symmetry as well as the spontaneous breaking of chiral symmetry in the
massless Thirring model does not contradict to Coleman's
theorem. Indeed, as has been discussed in \cite{FI2,FI4,FI7},
Coleman's theorem is applicable only to quantum field theories in
1+1--dimensional space--time with Wightman's observables defined on
the test functions $h(x)$ from ${\cal S}(\mathbb{R}^2)$, $h(x) \in
{\cal S}(\mathbb{R}^2)$, whereas the massless Thirring model and its
bosonized version are the quantum field theories with Wightman's
observables defined on the test functions $h(x)$ from ${\cal
S}_0(\mathbb{R}^2) = \{h(x) \in {\cal S}(\mathbb{R}^2); \tilde{h}(0) =
0\}$, where $\tilde{h}(0)$ is the Fourier transform of $h(x)$ at zero
momentum. Due to the vanishing of $\tilde{h}(0)$, the collective
zero--mode, responsible for infrared divergences of the free massless
(pseudo)scalar field theory bosonizing the massless Thirring model
\cite{FI2,FI6}, cannot be measured by Wightman's observables
\cite{FI2,FI4,FI7}.

\end{document}